\documentclass[aps, prb, superscriptaddress, reprint, nobibnotes]{revtex4-2}
\usepackage{amsmath,amssymb}
\usepackage{color}
\usepackage{comment}
\usepackage{bm,graphicx,hyperref}
\hypersetup{%
  breaklinks = {true},
  citecolor = {blue},
  colorlinks = {true},
  linkcolor = {red},}

\usepackage{lineno}

\begin{document}

\title{Single-piston quantum engine}
\author{A. Rodin}
\affiliation{Yale-NUS College, 16 College Avenue West, 138527, Singapore}
\affiliation{Centre for Advanced 2D Materials, National University of Singapore, 117546, Singapore}
\affiliation{Department of Materials Science and Engineering, National University of Singapore, 117575, Singapore}

\begin{abstract}

A single-piston quantum engine based on a harmonic oscillator acting as the working fluid is proposed.
Using the fact that the interaction between the piston and the oscillator depends on the extent of the oscillator wavefunction, one can control this interaction by modifying the oscillator temperature.
By retracting the piston when the interaction is weak (hot oscillator) and returning it to the original position when the coupling is strong (cold oscillator), useful work can be performed assuming the interaction is attractive.
The cycle of the engine is simulated numerically using two different powering protocols: bath and measurement.
Using the collision model for the baths, the engine is shown to reach a steady state with positive work output.

\end{abstract}	
\maketitle

\section{Introduction}
\label{sec:Introduction}

The purpose of an engine is to convert energy into work while operating cyclically.
Although essentially all commonplace implementations are rooted in classical physics for their operation, there is a growing interest in quantum engines.~\cite{Bhattacharjee2021, Myers2022}
Unlike their classical analogs, where the working fluid is typically a liquid or a gas, quantum engines employ quantum components.
Some examples of these quantum ``working fluids" are two-level systems~\citep{Huang2012, Chand2017a, Thomas2018, Chand2021}, single~\citep{DelCampo2014, Rossnagel2014, Kosloff2017, Reid2017, Cavaliere2022, Fei2022, Cavaliere2023, Rodin2024} or multiple~\citep{Boubakour2023, Carrega2023} harmonic oscillators, or photons~\citep{Zhang2014}.
Another feature that sets quantum and classical varieties apart is how they are powered.
In particular, classical implementations rely on hot reservoirs as energy sources and require cold baths to expel waste heat.
Quantum versions, on the other hand, can also obtain the required energy from measurements,~\citep{Chand2017, Elouard2017, Elouard2017a, Chand2018, Elouard2018, Das2019, Bresque2021, Manikandan2022, Alam2022, Jussiau2023} which act as an effective hot bath.

When discussing quantum engines, it is generally assumed that there is a way to transfer the work released during the power stroke of the cycle to some ``piston" without explicitly focusing on this engine component.
This paper proposes an implementation of a quantum engine in which the piston plays an integral role.
A recent study~\citep{Rodin2023} demonstrated that it is possible to modify the coupling strength between quantum engine components by controlling their energies.
In particular, heating two interacting harmonic oscillators increases the extent of their probability distributions, leading to a greater separation and weakening their interaction.
This temperature-dependent interaction strength is at the heart of the proposed cycle.

\begin{figure}
    \centering
    \includegraphics[width = \columnwidth]{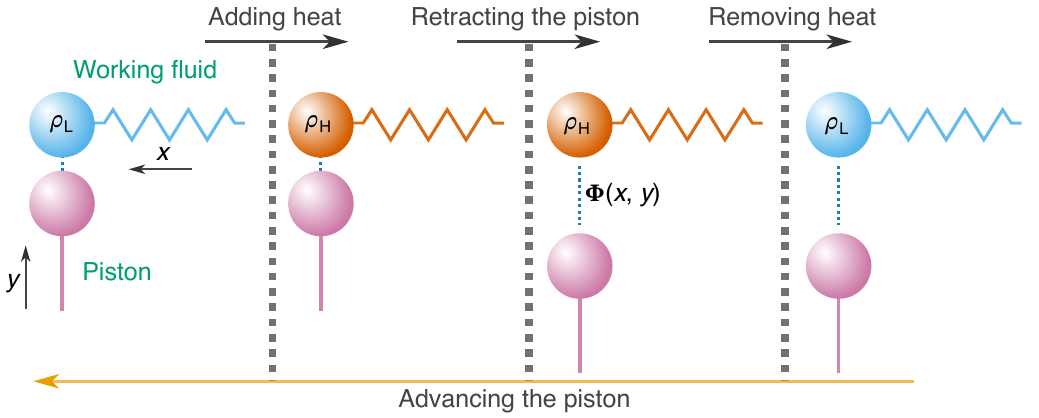}
    \caption{\emph{Engine cycle schematic.} The four phases of the engine cycle proceed in the direction indicated by the arrows. The working fluid is a quantum harmonic oscillator. It is coupled to an externally controlled piston via an attractive interaction, so adding heat to the working fluid reduces the coupling strength. Conversely, cooling the oscillator makes the interaction stronger. The power stroke of the cycle occurs when the piston is advanced and is denoted by the orange arrow.}
    \label{fig:Schematic}
\end{figure}

A cycle employing this energy-controlled coupling is shown in Fig.~\ref{fig:Schematic}.
The system consists of a harmonic oscillator, acting as the working fluid, and a moving piston.
The two components are coupled via an attractive interaction.
The cycle starts with the oscillator in a low-energy state and the piston positioned close to the oscillator's equilibrium point.
As the first step, the oscillator's energy is increased by adding heat to it.
This heating reduces the interaction strength between the oscillator and the piston due to an increased extent of the oscillator wavefunction, allowing the piston to be retracted from the oscillator in the second step of the cycle with a reduced energy cost.
With the piston retracted, the oscillator is cooled down using a cold bath.
Finally, the piston is returned to its initial position. 
Due to the enhanced interaction because of the narrower oscillator wavefunction in the final step, more work is released during this phase than was required to retract it, resulting in a net-positive work output.

As mentioned above, the heat required for quantum engine operation can originate either from heat baths or measurements.
Therefore, after introducing the general model for the engine in Sec.~\ref{sec:Model}, two prototypical approaches using a hot bath and measurements are demonstrated in Sec.~\ref{sec:Engine_operation}.
Summary and conclusions of the main results and the outlook for future research are found in Sec.~\ref{sec:Summary}.

All computations are performed using {\scshape julia}.~\citep{Bezanson2017}
The plots are made using Makie.jl package~\citep{Danisch2021} using the color scheme designed for colorblind readers.~\citep{Wong2011}
The scripts used for computing and plotting can be found at https://github.com/rodin-physics/quantum-oscillator-engine.

\section{Model}
\label{sec:Model}

As shown in Fig.~\ref{fig:Schematic}, the engine consists of a harmonic oscillator, acting as the working fluid, coupled to an externally controlled piston.
When the working fluid is decoupled from the environment, the corresponding time-dependent Hamiltonian, expressed in terms of the oscillator energy $\hbar\Omega$, is given by

\begin{equation}
    \hat{H}(\tau) =  \left(\hat{a}^\dagger \hat{a} + \frac{1}{2}\right) + \Phi\left[\hat{x},y(\tau)\right]\,.
    \label{eqn:H}
\end{equation}
Here, the first term describes an independent oscillator using the second-quantization operators and $\Phi$ is the interaction between the two engine components, where $\hat{x}$ is the oscillator's position operator and $y(\tau)$ is the piston coordinate.
In addition to using $\hbar\Omega$ as the energy scale, it is convenient to express lengths in terms of the quantum oscillator length and time in terms of the oscillator periods so that $t = 2\pi \tau / \Omega$.

For this study, the interaction is set to

\begin{equation}
    \Phi\left[\hat{x},y(\tau)\right] = \Phi_{0}\exp\left[-\frac{\hat{x}^2 + y^2(\tau)}{2\sigma^2}\right]\,.
    \label{eqn:Phi}
\end{equation}
There are two reasons for choosing this form of $\Phi$.
First, the amplitude and the extent of the Gaussian are easily tunable, making this type of interaction very convenient for illustrating the relevant behavior.
Second, $y$ and $\hat{x}$ are separable, simplifying the computational procedure.
With this choice, the matrix elements of $\hat{H}(\tau)$ in the Fock space become

\begin{equation}
    \langle j|\hat{H}(\tau)|k\rangle = \left(j+\frac{1}{2}\right)\delta_{j,k} + \Phi_0^{-\frac{ y^2(\tau)}{2\sigma^2}}
     \langle j|e^{-\frac{\hat{x}^2}{2\sigma^2}}|k\rangle\,.
     \label{eqn:Matrix_elements}
\end{equation}
The form of Eq.~\eqref{eqn:Matrix_elements} indicates that the interaction matrix must be computed only once for a particular choice of $\sigma$ and scaled based on $y(\tau)$ as the piston moves.
Although $\langle j|e^{-\frac{\hat{x}^2}{2\sigma^2}}|k\rangle$ can be calculated analytically for the quantum oscillator wavefunctions, producing Gaussian hypergeometric functions, the integrals in the simulations are taken using Gaussian quadratures to avoid potential issues associated with the numerical implementation of these special functions.
For simplicity, the piston is assumed to move at a constant speed so that $y(\tau) = y_\mathrm{init} + \tau(y_\mathrm{final} - y_\mathrm{init}) / \tau_p $ for $0\leq \tau \leq \tau_p$, where $\tau_p$ is the duration of the retraction and advancing phases.

\begin{figure}
    \centering
    \includegraphics[width = \columnwidth]{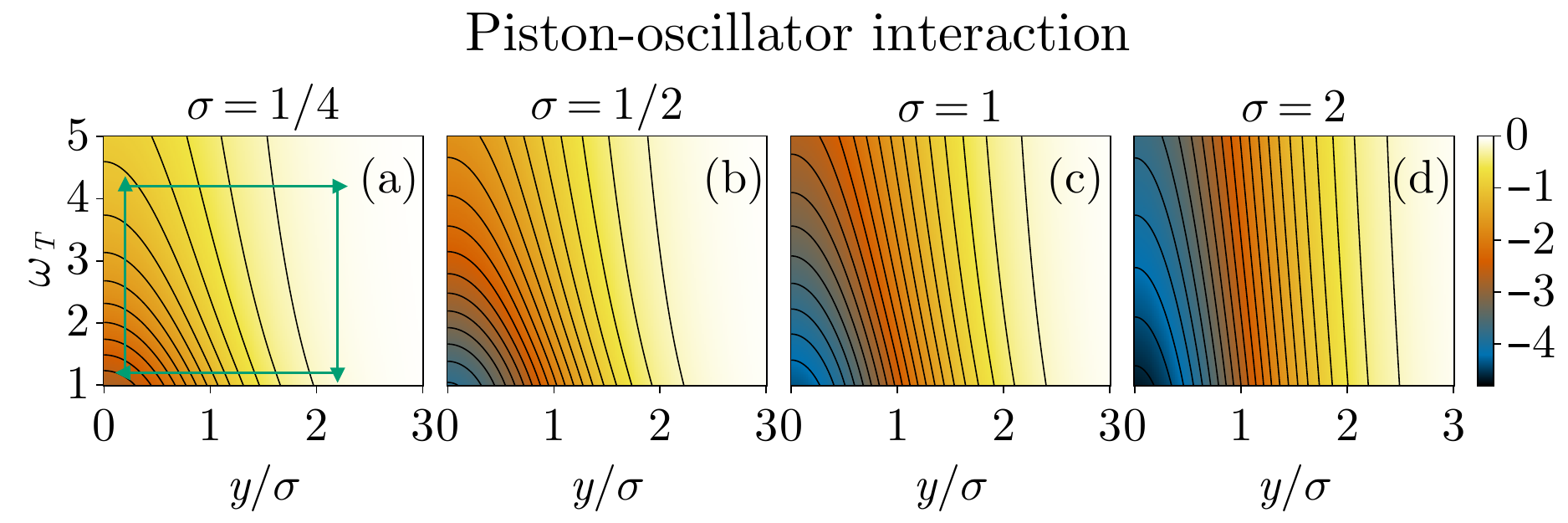}
    \caption{\emph{Controlling the interaction.} Dependence of the interaction energy on temperature and piston position for $\Phi_0 = -5$ and several values of $\sigma$. The black lines are energy equicontours separated by $0.2$. Retracting the piston or raising the oscillator temperature reduces the interaction strength. The green contour indicates an idealized engine cycle with vertical lines denoting heat transfer and the horizontal ones piston movement. The energy added or removed during a given phase is proportional to the number of equicontour lines crossed by the corresponding cycle segment.}
    \label{fig:Piston_interaction}
\end{figure}

With the interaction form chosen, it is now possible to demonstrate the key component of the engine operation: the temperature-dependent coupling between the working fluid and the piston.
For this illustration, the oscillator is assumed to be in a thermal state, described by the density operator $\hat{\rho} = e^{-\hat{H} / \omega_T}/ \mathrm{tr}\left(e^{-\hat{H} / \omega_T}\right)$, where $\omega_T = k_B T / \hbar\Omega$ is the thermal frequency corresponding to temperature $T$ and $k_B$ is the Boltzmann constant.
The interaction energy is then given by $\mathrm{tr}\left(\hat{\Phi}\hat{\rho}\right)$ with $\hat{\Phi}$ matrix elements given by the second term of Eq.~\eqref{eqn:Matrix_elements}.
Setting $\Phi_0 = -5$, the value that will be used in subsequent simulations, the energy is calculated as a function of $\omega_T$ and $y$ for several values of $\sigma$ with the results given in Fig.~\ref{fig:Piston_interaction}.
As expected, the system's energy is the lowest for a wide interaction term (large $\sigma$) when the piston is close to the oscillator and the temperature is low.
Increasing either $\omega_T$ or $y$ reduces the magnitude of the interaction.

Figure~\ref{fig:Piston_interaction} provides a convenient way of thinking about the cycle described by the schematic in Fig.~\ref{fig:Schematic}.
Let the state of the system be given by a point in the $\omega_T$-$y/\sigma$ plane, indicating that the working fluid is in a thermal state.
Following Fig.~\ref{fig:Schematic}, one starts at a small $y$ and $\omega_T$ and, by adding the heat to the system and assuming that the oscillator state remains in a thermal state, moves the point vertically.
Next, with $\omega_T$ fixed, the piston is retracted and the point moves horizontally to the right.
As the heat is removed from the system, the point moves down.
Finally, the piston is advanced and the point returns to its starting position.
An example of such a path is given in Fig.~\ref{fig:Piston_interaction}(a).
If the piston is moved adiabatically, the state of the oscillator does not change and only the interaction energy is altered so that the amount of energy added to or removed from the system during these phases is proportional to the number of equicontours crossed by the path.
Therefore, as long as more equicontours are crossed by the left-moving segment than the right-moving one, the engine outputs work.

It is useful to provide an estimate for the efficiency associated with this engine, given by $\eta = (W_\mathrm{out} - W_\mathrm{in})/Q_\mathrm{added}$, where $Q_\mathrm{added}$ corresponds to the heat added to the system during the first phase and $W$ is the work extracted from or added to the system during the second and the fourth phases.
Let $\rho_{H/L}$ correspond to the oscillator's density operator in its high/low energy state, $\Phi_\mathrm{close}$ and $\Phi_\mathrm{far}$ give the interaction when the piston is close to/far from the oscillator, and $H_0$ be the first term in Eq.~\eqref{eqn:H}.
Assuming that the piston motion is adiabatic so that $\rho_{H/L}$ remains fixed, $\eta$ becomes

\begin{align}
    \eta & = 
    \frac{\mathrm{tr}\left[\rho_H\left(\Phi_\mathrm{close} - \Phi_\mathrm{far}\right)\right]-\mathrm{tr}\left[\rho_L\left(\Phi_\mathrm{close} - \Phi_\mathrm{far}\right)\right]}{\mathrm{tr}\left[\rho_H\left(H_0 + \Phi_\mathrm{close}\right)\right]-\mathrm{tr}\left[\rho_L\left(H_0 + \Phi_\mathrm{close}\right)\right]}
     \nonumber
     \\
          &=   \frac{
     \mathrm{tr}\left[\left(\rho_H-\rho_L\right)\Phi_\mathrm{close}\right]-\mathrm{tr}\left[\left(\rho_H-\rho_L\right)\Phi_\mathrm{far}\right]}{
     \mathrm{tr}\left[\left(\rho_H-\rho_L\right)H_0\right] + \mathrm{tr}\left[\left(\rho_H-\rho_L\right)\Phi_\mathrm{close}\right]}
     \nonumber
     \\
     &\approx  \frac{1}{\frac{\mathrm{tr}\left[\left(\rho_H-\rho_L\right) H_0\right]}{\mathrm{tr}\left[\left(\rho_H-\rho_L\right) \Phi_\mathrm{close}\right]}+1}\,,
     \label{eqn:eta}
\end{align}
where the approximation in the final line holds when $\Phi_\mathrm{far}\rightarrow 0$.

Equation~\eqref{eqn:eta} indicates that, for a particular oscillator (fixed $H_0$), the efficiency is maximized by increasing $\mathrm{tr}\left[\left(\rho_H-\rho_L\right) \Phi_\mathrm{close}\right]$.
Using the diagram in Fig.~\ref{fig:Piston_interaction}(a), this requirement means that one should maximize the number of equicontour lines that the upwards-moving segment crosses.
Thus, from Fig.~\ref{fig:Piston_interaction}, for the parameters chosen, $\sigma = 2$ is too wide as most of the energy delivered to the system comes from piston retraction rather than from heating.
On the other hand, $\sigma = 1/4$ is inferior to $\sigma = 1/2$, where the equicontours are denser.
This non-monotonic dependence of $\eta$ on $\sigma$ suggestes that one can improve the efficiency by exploring different interaction profiles not limited to a Gaussian form.

To estimate $\eta$, the piston movement was taken to be adiabatic.
If it is not, the energy is added to the oscillator in a thermal state during the piston movement and has to be either removed as waste heat (after the piston retraction) or results in a reduced work output (during the piston advance).
Of course, for the engine to deliver a finite power, the individual phases have to take a finite amount of time.
Therefore, it is useful to explore how slow is ``slow enough" when it comes to the piston motion using the following procedure.

Starting with the Hamiltionians corresponding to retracted ($y = 10\sigma$) and advanced ($y = 0$) piston, denoted by $\hat{H}_R$ and $\hat{H}_A$, respectively, one determines their ground states $|R\rangle$ and $|A\rangle$.
If the piston were retracted (advanced) adiabatically, the energy of the system would end up as $\langle R |\hat{H}_R|R\rangle$ ($\langle A |\hat{H}_A|A\rangle$) because the system would remain in its ground state.
Conversely, for an instantaneous change of piston position, the state would not have time to evolve and the final energy would be $\langle A |\hat{H}_R|A\rangle$ ($\langle R |\hat{H}_A|R\rangle$).
Hence, the actual final energy is bounded by these two values.

\begin{figure}
    \centering
    \includegraphics[width = \columnwidth]{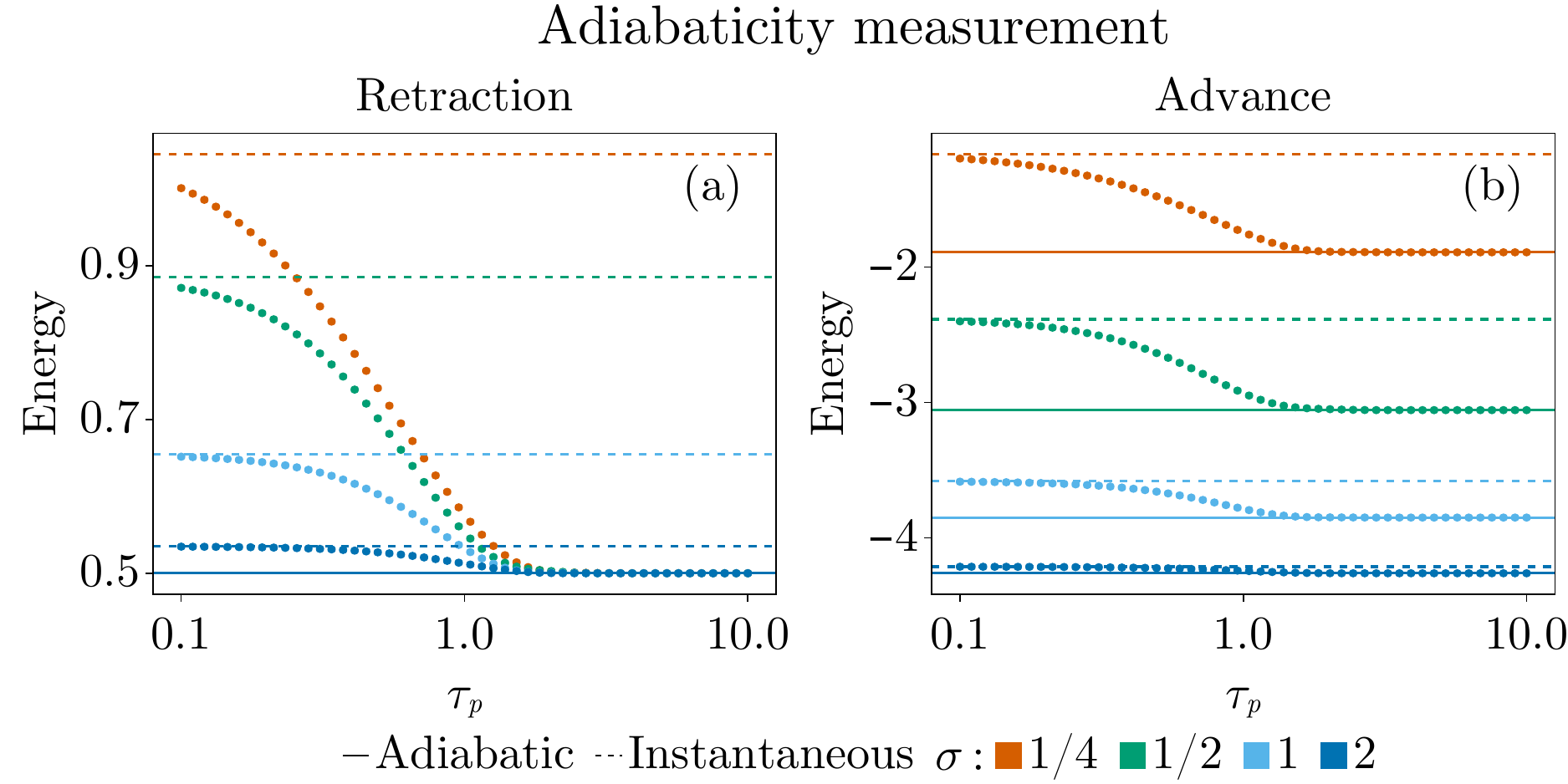}
    \caption{\emph{Role of finite $\tau_p$.} System energy as a function of time of piston retraction and advance $\tau_p$ for $\Phi_0 = -5$ and several values of $\sigma$. The piston moves between $y = 0$ and $y = 10\sigma$. The numerically-computed results for finite $\tau_p$ are bounded by the adiabatic ($\tau_p\rightarrow\infty$) and instantaneous ($\tau_p\rightarrow 0$) results. Smaller $\sigma$ leads to greater non-adiabatic effects.}
    \label{fig:Piston_movement}
\end{figure}

To obtain the system energy for a finite $\tau_p$, one uses the fact that the state evolves following the time-dependent Schr\"{o}dinger equation

\begin{equation}
    \frac{d}{d\tau}|\Psi(\tau)\rangle = -2\pi i \hat{H}(\tau) |\Psi(\tau)\rangle\,,
    \label{eqn:Psi_diff_eq}
\end{equation}
where the factor of $2\pi$ originates from the definition of $\tau$.
Solving  Eq.~\eqref{eqn:Psi_diff_eq} using the fifth order Runge-Kutta method starting with $|\Psi(0)\rangle = |A\rangle$ and $|\Psi(0)\rangle = |R\rangle$, the final energy can be computed as a function of $\tau_p$ for a constant-speed piston with the results given in Fig.~\ref{fig:Piston_movement}.
As expected, the finite-$\tau_p$ energies lie between the adiabatic and instantaneous results for both retraction and advance of the piston.
Figure~\ref{fig:Piston_movement} suggests that $\tau_p \gtrapprox 5$ is sufficiently slow to avoid the non-adiabatic effects associated with the piston movement.

\section{Engine operation}
\label{sec:Engine_operation}

\subsection{Operation protocol}
\label{sec:Operation_protocol}

Conceptually, the most straightforward way to heat up or cool down an oscillator mode is to connect it to a thermal bath.
For the sake of illustration, this paper models the baths using single oscillator modes described by a density operator corresponding to a thermal state, as was done in Ref.~\citep{Rodin2024}.
Every time the working fluid is connected to a bath, the latter is reset to the appropriate thermal state.
The process of resetting a relevant component to some predetermined configuration can be regarded as having the working fluid interact with a series of identical states, a scheme known as the collision model.~\citep{Ciccarello2022}
Naturally, using a single mode as a bath will not bring the working fluid to a thermal state with the bath temperature.~\citep{Rodin2024}
Therefore, the efficiency formula in Eq.~\eqref{eqn:eta} is not expected to hold.
Nevertheless, as will be shown, the correct direction of the energy flow between the working fluid and the baths will be sufficient to demonstrate the engine operation.

To keep the discussion as streamlined as possible, the frequency of the bath oscillators is taken to be identical to that of the working fluid.
In addition to eliminating a parameter, setting the oscillator frequencies to the same value facilitates the coupling between the modes.
The bath coupling is Gaussian, just as the piston interaction term, given by

\begin{equation}
    Y(\hat{x},\hat{z}) = Y_0 \exp\left[-\frac{(\hat{x} - x_0)^2 + (\hat{z} - z_0)^2}{2\lambda^2}\right]\,,
    \label{eqn:Bath_Coupling}
\end{equation}
where $\hat{z}$ is the position of the bath oscillator.
The offsets $x_0$ and $z_0$ mean that the equilibrium points of the two oscillators do not coincide in the $xz$ plane and are introduced to allow modes with different parities to couple, which would be forbidden by a symmetric potential $e^{-\hat{x}^2 / 2\lambda^2}$.
Setting $x_0 = z_0 = \lambda = 1$ and using the separability of the interaction term, one can write down the full interaction matrix $Y = Y_0 Y_\mathrm{single}\otimes Y_\mathrm{single}$, where $Y_\mathrm{single}$ elements are given by with elements $\langle j |e^{-(x - x_0)^2/2\lambda^2}|k\rangle$ for all the Fock states in the single-oscillator basis.
Just as the matrix elements in Eq.~\eqref{eqn:Matrix_elements}, these are computed using Guassian quadratures.

For simplicity, the interaction between the bath and the working fluid is assumed to be switched on and off instantaneously so that, when the piston is stationary, the Hamiltonian is time-independent, leading to

\begin{align}
    \hat{H}_\mathrm{bath} &=  \left[\left(\hat{a}^\dagger \hat{a} + \frac{1}{2}\right) + \Phi\left(\hat{x},y\right)\right]\otimes \mathbf{1}
    \nonumber
    \\
    &+  \mathbf{1}\otimes\left(\hat{b}^\dagger \hat{b} + \frac{1}{2}\right)
    + Y_0 Y_\mathrm{single}\otimes Y_\mathrm{single}\,,
    \label{eqn:H_heat}
\end{align}
where $\mathbf{1}$ is the identity.
The corresponding time evolution operator $\hat{\mathcal{B}} = \exp\left(-2\pi i \tau_b \hat{H}_\mathrm{bath}\right)$, where $\tau_b$ is the contact time between the fluid and the bath, taken to be the same for both hot and cold baths.
The baths' thermal states are given by $\hat{\rho}_b = \exp\left(-b^\dagger b / \omega_T\right)/\mathrm{tr}[\exp\left(-b^\dagger b / \omega_T\right)]$, where $\omega_T$ is the bath temperature.

To move the piston, instead of working with the oscillator state as was done in Sec.~\ref{sec:Model} when studying the role of $\tau_p$, it is better to calculate the time evolution operator from 

\begin{equation}
    \frac{d}{d\tau}\hat{\mathcal{U}}(\tau, \tau') = -2\pi i \hat{H}(\tau) \hat{\mathcal{U}}(\tau, \tau')\,,
    \label{eqn:U_diff_eq}
\end{equation}
which is solved using the fifth order Runge-Kutta method with $\hat{\mathcal{U}}(\tau',\tau') = 1$.
To understand the advantage of $\hat{\mathcal{U}}$, it is illustrative to consider a particular implementation, starting with the bath-powered setup.

\subsection{Bath-powered setup}
\label{sec:Bath_powered_setup}

Following Fig.~\ref{fig:Schematic}, the cycle begins with heat addition.
Thus, if the state of the working fluid in the beginning of the $n$th cycle is given by $\hat{\rho}_n$, the state of the fluid in the beginning of the following cycle is

\begin{align}
    \hat{\rho}_{n+1} =\hat{\mathcal{U}}^\dagger \mathrm{tr}_b \left[\hat{\mathcal{B}} \left\{\hat{\mathcal{U}}\mathrm{tr}_b\left[\hat{\mathcal{B}}\left(\hat{\rho}_n\otimes \hat{\rho}_h\right)\hat{\mathcal{B}}^\dagger\right]\hat{\mathcal{U}^\dagger}\otimes \hat{\rho}_c\right\}\hat{\mathcal{B}}^\dagger\right] \hat{\mathcal{U}}\,,
    \label{eqn:Bath_Cycle}
\end{align}
where $\hat{\mathcal{U}}$ corresponds to the operator describing the piston retraction.
Equation~\eqref{eqn:Bath_Cycle} should be read from the inside outward to follow the cycle.
First, the fluid is coupled to the hot bath in thermal state $\hat{\rho}_h$, as shown by the tensor product.
After that, the composite system is allowed to evolve in time by applying operators $\hat{\mathcal{B}}$ and $\hat{\mathcal{B}}^\dagger$.
To decouple the working fluid from the bath, a partial trace $\mathrm{tr}_b$ is performed with respect to the bath.
Next, the piston is retracted by sandwiching the fluid state between $\hat{\mathcal{U}}$ and $\hat{\mathcal{U}}^\dagger$.
Then, the fluid is coupled to a cold bath $\hat{\rho}_c$ and the two-oscillator system is evolved using the same operator $\hat{\mathcal{B}}$ as was used for the hot bath, followed by a decoupling.
Finally, the piston is advanced, as can be seen from the reversed application of $\hat{\mathcal{U}}^\dagger$ and $\hat{\mathcal{U}}$, completing the cycle.
Thus, instead of evolving the state every time the piston moves, one needs to compute $\hat{\mathcal{U}}$ only once making a multi-cycle calculation more efficient.

\begin{figure}
    \centering
    \includegraphics[width = \columnwidth]{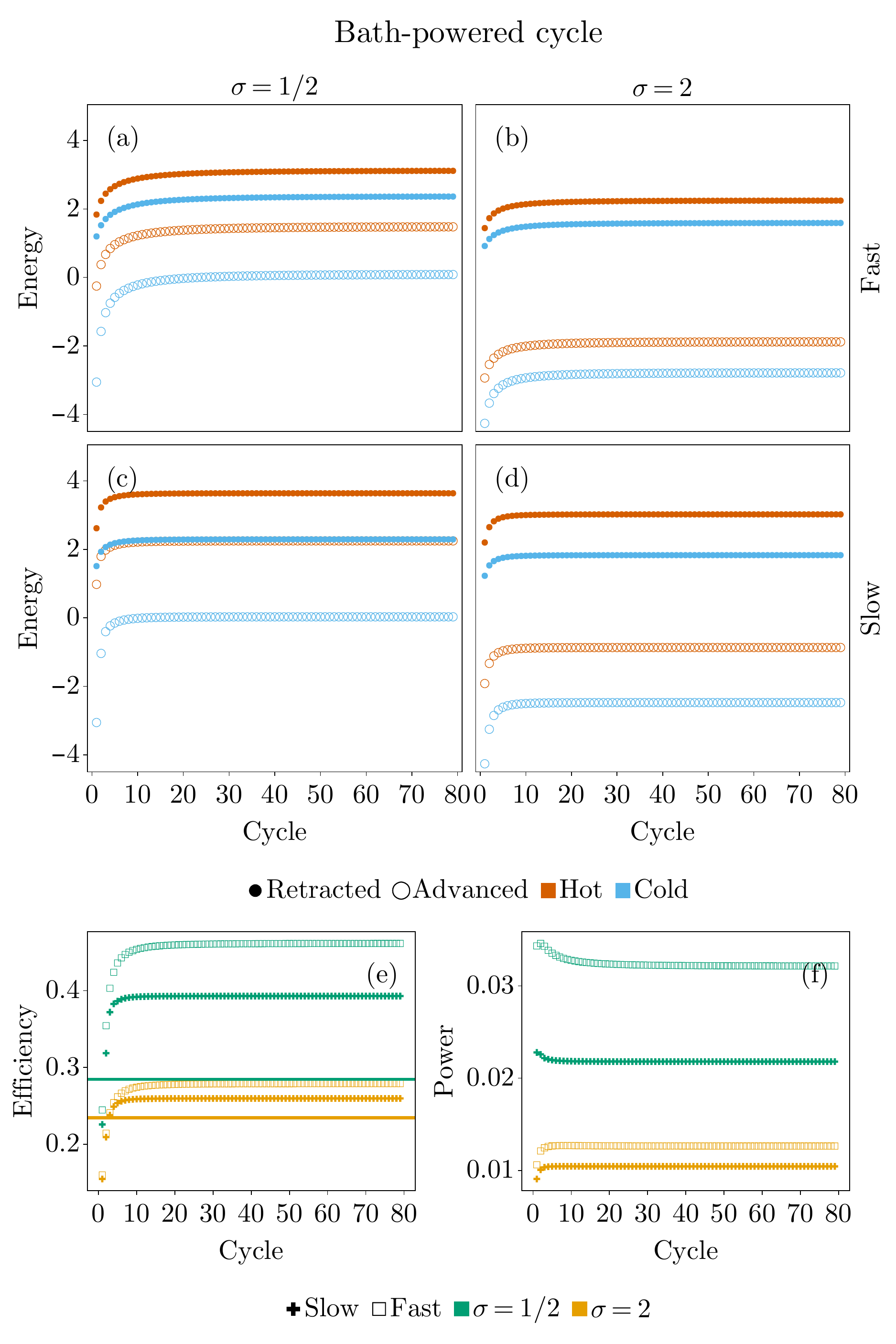}
    \caption{\emph{Bath engine operation}. 80 cycles of a bath-powered engine initialized in a thermal state and evolved using Eq.~\eqref{eqn:Bath_Cycle} with $\omega_T^\mathrm{hot} = 5$ and $\omega_T^\mathrm{cold} = 1/10$.
    The interaction between the baths and the gas is given by Eq.~\eqref{eqn:Bath_Coupling} with $Y_0 = x_0 = z_0 = 1$. 
    The baths and the oscillator Fock spaces contain 51 states.
    The piston interaction is given by Eq.~\eqref{eqn:Phi} with $\Phi_0 = -5$.\\
    (a)-(d) System energy at each phase of the cycle as a function of cycle number. For the fast engine (top row), the time of each stroke is 5, while for the slow one (bottom row) it is 10. (e) Efficiency for each of the engines as a function of the cycle number, computed by dividing the total work output by heat input. (f) The power of the engines, obtained by dividing the total work output by the duration of a cycle.}
    \label{fig:Bath_engine}
\end{figure}

To demonstrate the engine operation, the four phases of the cycle are assumed to have the same duration $\tau_p$.
Two different $\sigma$'s ($1/2$ and $2$) and $\tau_p$'s (5 and 10) are used for a total of four configurations.
The bath temperatures are set to $\omega_T^\mathrm{cold} = 1/10$ and $\omega_T^\mathrm{hot} = 5$, and the Fock basis for each oscillator contains 51 states.
For each configuration, the working fluid is initialized in the thermal state at $\omega_T^\mathrm{cold}$ with the piston in the advanced position.
It is taken 80 times through the cycle described by Eq.~\eqref{eqn:Bath_Cycle}.
At the end of each stroke, the energy of the working fluid is computed by taking the trace of the product of its density operator with $\left(\hat{a}^\dagger \hat{a} + \frac{1}{2}\right) + \Phi\left(\hat{x},y\right)$ for the appropriate value of $y$.
The computed energies for the four realizations are given in Fig.~\ref{fig:Bath_engine}(a)-(d).
The $x$-coordinate labels the cycle and, for each cycle, the order of the points is ``advanced cold" $\rightarrow$ ``advanced hot" $\rightarrow$ ``retracted hot" $\rightarrow$ ``retracted cold," after which one moves to ``advanced cold" of the next cycle.

Figure~\ref{fig:Bath_engine} demonstrates that even the single-oscillator bath, where the working fluid does not actually reach a thermal state, is sufficient for the engine to output work as it reaches a steady state.
One can see that the faster cycle shows a small creep in energy, more substantial for $\sigma = 1/2$.
This creep can be attributed to the non-adiabatic effects of the piston movement: according to Fig.~\ref{fig:Piston_movement}, smaller $\sigma$'s are more susceptible to this effect.
In an experimental setup, this parasitic behavior can be mitigated by having the working fluid interact with a real baths which are able to absorb more heat from the oscillator than a single mode used as a bath here.

Figure~\ref{fig:Bath_engine}(e) plots the cycle-resolved efficiency for each setup, shown to also reach a steady state.
Theoretical efficiency obtained from Eq.~\eqref{eqn:eta}, where $\rho_{H/L}$ are taken to be thermal states with $\omega_T^{\mathrm{hot}/\mathrm{cold}}$, is plotted along with the simulation results. 
Because the working fluid does not reach a thermal state, the result does not agree quantitatively with the numerics.
Nevertheless, as discussed after Eq.~\eqref{eqn:eta}, $\sigma = 1/2$ delivers a better efficiency than $\sigma = 2$ for both theoretical and numerical results.

Curiously, the efficiency of the fast cycles is higher than the slow ones.
This efficiency gain, however, comes at the expense of the work performed.
Figure~\ref{fig:Bath_engine}(f) provides the cycle-resolved power for each engine configuration obtained by dividing the net work output by $4\tau_p$.
Even though the fast cycle takes half the time compared to the slow one, its power is not doubled, indicating that the slow cycle delivers more work.
One can confirm this statement by comparing the separation between red markers for each cycle to that of the blue markers.
The former corresponds to the work done on the engine, while the latter is the work done by the engine.
The greater the difference, the more net work the engine outputs.
It is evident that the difference is larger for the slow cycles for both $\sigma$'s.

\begin{figure}
    \centering
    \includegraphics[width = \columnwidth]{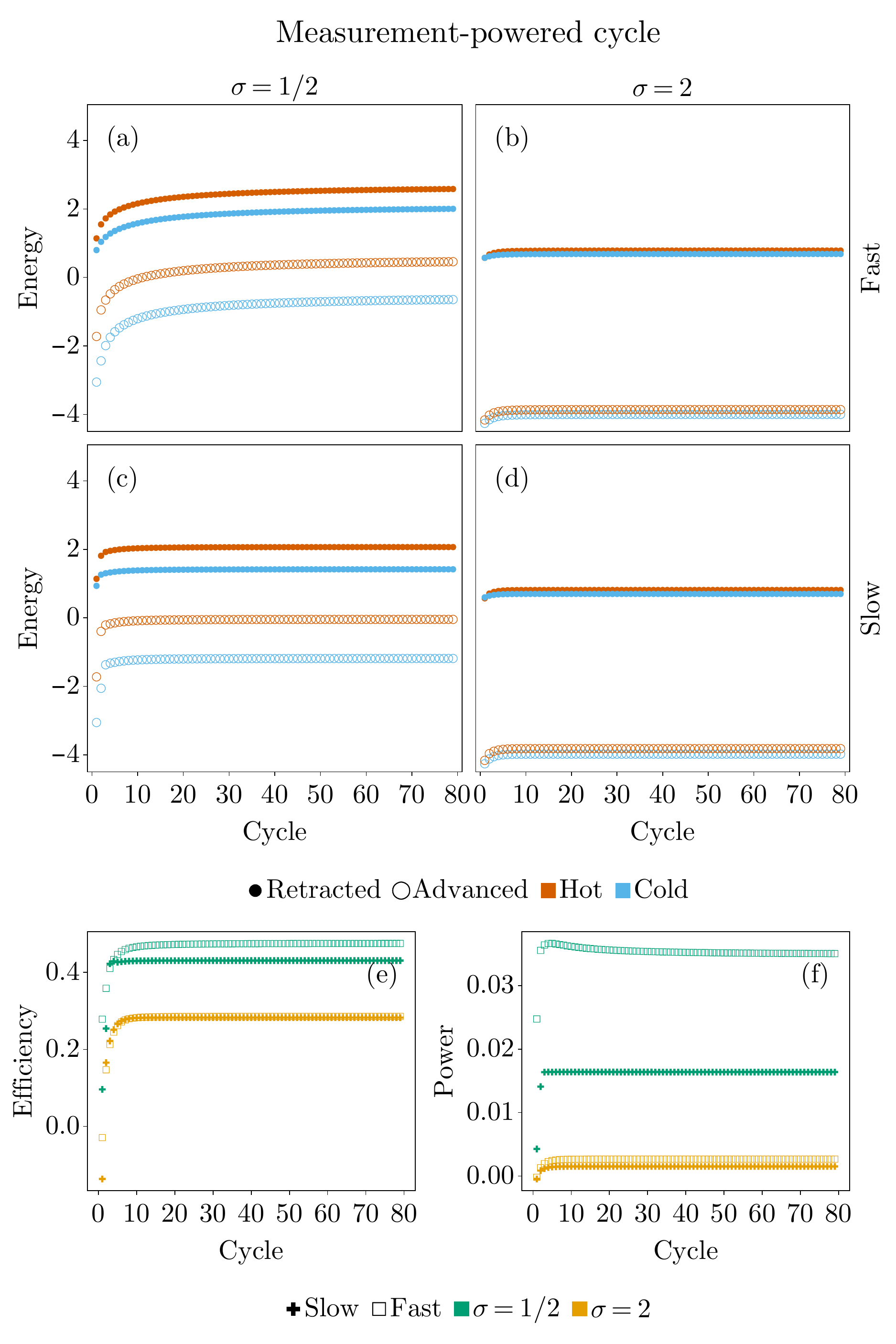}
    \caption{\emph{Measurement engine operation}. 80 cycles of a measurement-powered engine initialized in a thermal state and evolved using Eq.~\eqref{eqn:Measurement_Cycle} with $\omega_T^\mathrm{cold} = 1/10$.
    The interaction between the baths and the gas is given by Eq.~\eqref{eqn:Bath_Coupling} with $Y_0 = x_0 = z_0 = 1$. 
    The baths and the oscillator Fock spaces contain 51 states.
    The piston interaction is given by Eq.~\eqref{eqn:Phi} with $\Phi_0 = -5$.\\
    (a)-(d) System energy at each phase of the cycle as a function of cycle number. For the fast engine (top row), the time of each stroke is 5, while for the slow one (bottom row) it is 10. (e) Efficiency for each of the engines as a function of the cycle number, computed by dividing the total work output by heat input. (f) The power of the engines, obtained by dividing the total work output by the duration of a cycle.
    }
    \label{fig:Measurement_Engine}
\end{figure}

\subsection{Measurement-powered setup}

Changing to power source from a hot bath to measurements while keeping everything else the same amounts to replacing $\mathrm{tr}_b\left[\hat{\mathcal{B}}\left(\hat{\rho}_n\otimes \hat{\rho}_h\right)\hat{\mathcal{B}}^\dagger\right] \rightarrow \mathrm{diag}\left(\hat{\rho}_n\right)$ in Eq.~\eqref{eqn:Bath_Cycle}, corresponding to the measurement.
$\mathrm{diag}\left(\hat{\rho}_n\right)$ means that the off-diagonal elements (coherence terms) are set to zero, giving a classical probability.
Hence, the cycle expression becomes

\begin{align}
    \hat{\rho}_{n+1} =\hat{\mathcal{U}}^\dagger \mathrm{tr}_b \left\{\hat{\mathcal{B}} \left[\hat{\mathcal{U}}\,\mathrm{diag}\left(\hat{\rho}_n\right)\hat{\mathcal{U}^\dagger}\otimes \hat{\rho}_c\right]\hat{\mathcal{B}}^\dagger\right\} \hat{\mathcal{U}}\,.
    \label{eqn:Measurement_Cycle}
\end{align}

The results for a set of four simulations using the same two values of $\sigma$ and phase duration as above are given in Fig.~\ref{fig:Measurement_Engine}.
While the $\sigma = 1/2$ setup looks qualitatively similar to the corresponding configurations in Fig.~\ref{fig:Bath_engine}, including the energy creep in the fast cycle, $\sigma = 2$ is drastically different, showing virtually no work output.
This outcome is attributable to the fact that the piston-generated potential is wide on the scale of the the oscillator's wavefunctions, resulting in a very weak harmonic mixing.
Therefore, the measurement does not result in a substantial transfer of energy to the oscillator since the oscillator does not transition to higher energy states.

The efficiency for the measurement-powered cycle is similar to the bath-powered one, as seen by comparing Figs.~\ref{fig:Bath_engine}(e) and \ref{fig:Measurement_Engine}(e).
To compute the power, the net work output is divided by $3\tau_p$ since the measurement is assumed to be an instantaneous process.
The power for $\sigma = 1/2$ is comparable to the bath-powered setup.
In the $\sigma = 2$ case, on the other hand, the vanishing work output results in negligible power.

The key takeaway of this section is that, despite the restricted and artificial form of the baths, the engine reaches a steady state and is able to output useful work.
Enhancing the oscillator's ability to expel the waste heat by using a real reservoir and optimizing the piston and bath interaction profiles will improve the work output.

\section{Summary}
\label{sec:Summary}

This work has introduced and simulated a realization of a single-piston quantum engine, where the role of the working fluid is played by a harmonic oscillator.
By taking advantage of the fact that the interaction between the working fluid and the piston can be controlled by modifying the oscillator's energy, it has been shown that by following a cycle comprised of heating/cooling of the oscillator and piston motion towards and away from the oscillator, the engine can output work.
Two general protocols of engine fueling have been discussed: bath-powered and measurement-powered with both successfully demonstrating stable work output over multiple cycles, indicating a steady state of operation.
For simplicity, the heat reservoirs used here comprised of single thermal harmonic oscillator modes.
Even though these modes do not function as true thermodynamic baths, the fact that the engine produces work suggests that the engine operation scheme is robust and will perform better in the presence of real thermodynamic reservoirs.

There are several research directions that naturally flow from this study.
For an experimental realization of this system, the harmonic oscillator component could be implemented using cold atoms in an optical trap.
To extract the waste heat, one could employ laser cooling.
Therefore, a theoretical study could explore the operation of the proposed engine where the role of the working fluid is played by a harmonic oscillator with a controllable damping.
In a similar vein, powering the engine using resonant driving is an option. Naturally, when the piston is in the advanced position, the oscillator Fock states are no longer the eigenstates, simply driving at the natural oscillator frequency might not be optimal in every case.
Finally, a two-piston setup could be explored, where some of the waste heat expelled by one of the ``cylinders" goes to heating the other one.

\acknowledgements

The author acknowledges the National Research Foundation, Prime Minister Office, Singapore, under its Medium Sized Centre Programme and the support by Yale-NUS College (through Start-up Grant).


%

\end{document}